\pdfoutput=1
\documentclass[a4paper, 10pt, conference]{ieeeconf}
\IEEEoverridecommandlockouts
\usepackage{cite}
\usepackage{acronym}
\usepackage{amsmath,amssymb,amsfonts}
\usepackage{algorithmic}
\usepackage{graphicx}
\usepackage{textcomp}
\usepackage{url}
\usepackage{xcolor}
\usepackage{color}
\usepackage{booktabs}
\usepackage{tikz}
\usepackage{gensymb}
\usepackage{dblfloatfix}
\usepackage{placeins}

\usetikzlibrary{shapes.geometric, arrows.meta}
\usetikzlibrary{arrows.meta,positioning}
\definecolor{boxgray}{HTML}{EAEEF3}
\definecolor{boxteal}{HTML}{1CA3AD}
\definecolor{boxdteal}{HTML}{2E6E8E}
\definecolor{boxnavy}{HTML}{16243F}
\definecolor{elicg}{HTML}{DCEFE0}
\definecolor{elicgs}{HTML}{6FAE7C}
\definecolor{salb}{HTML}{DBE9FB}
\definecolor{salbs}{HTML}{7FA8DE}
\definecolor{ctxg}{HTML}{EEECE7}
\definecolor{ctxgs}{HTML}{9B9689}
\definecolor{ioc}{HTML}{FDF6EC}
\definecolor{iocs}{HTML}{C9A24A}
\graphicspath{{./figures/}{./}}
\def\BibTeX{{\rm B\kern-.05em{\sc i\kern-.025em b}\kern-.08em
    T\kern-.1667em\lower.7ex\hbox{E}\kern-.125emX}}

\usepackage[capitalize]{cleveref}
\crefname{section}{Section}{Sections}
\Crefname{section}{Section}{Sections}
\Crefname{table}{Table}{Tables}
\crefname{table}{Table}{Tables}
\crefname{figure}{Figure}{Figures}
\Crefname{figure}{Figure}{Figures}
\crefname{equation}{Eq.}{Eqs.}
\Crefname{equation}{Eq.}{Eqs.}

\acrodef{HMD}{head-mounted display}
\acrodef{ERP}{equirectangular projection}
\acrodef{QP}{quantization parameter}
\acrodef{CC}{correlation coefficient}
\acrodef{NSS}{normalized scanpath saliency}
\acrodef{KLD}{Kullback--Leibler divergence}
\acrodef{AUCJ}[AUC-J]{area under the ROC curve, Judd variant}
\acrodef{ODI}{omnidirectional image}
\acrodef{GT}{ground truth}
\acrodef{CTU}{coding tree unit}
\acrodef{RDO}{rate-distortion optimization}
\acrodef{ROI}{region of interest}
\acrodef{CNN}{convolutional neural network}
\acrodef{GAN}{generative adversarial network}
\acrodef{MSE}{mean squared error}
\acrodef{WSPSNR}[WS-PSNR]{weighted-to-spherically-uniform PSNR}
\acrodef{ELIC}{Efficient Learned Image Compression}
\acrodef{QVRF}{Quantization-error-aware Variable Rate Framework}
\acrodef{BDRATE}[BD-rate]{Bj{\o}ntegaard delta rate}
\acrodef{BDPSNR}[BD-PSNR]{Bj{\o}ntegaard delta PSNR}
\acrodef{WSSALPSNR}[WS-SAL-PSNR]{weighted-to-spherically-uniform saliency PSNR}

\begin{document}

\title{\LARGE \bf How Much Does Saliency Accuracy Matter for\\Saliency-Guided 360-degree Image Compression?}

\author{Sakib Ahmed$^{1}$, Sidi-Mohammed Senouci$^{1}$, Philippe Brunet$^{1}$,
Guillaume Caron$^{2,3}$, and Nevrez Imamoglu$^{3,4}$%
\thanks{\copyright~2027 IEEE. Personal use of this material is permitted.
Permission from IEEE must be obtained for all other uses, in any current or
future media, including reprinting/republishing this material for advertising
or promotional purposes, creating new collective works, for resale or
redistribution to servers or lists, or reuse of any copyrighted component of
this work in other works. This is the author's accepted version of a paper
accepted for publication at the 2027 IEEE/SICE International Symposium on
System Integration (SII 2027).}%
\thanks{$^{1}$Universit\'e Bourgogne Europe, DRIVE UR 1859, 58000 Nevers, France\newline
{\tt\small {firstname.surname}@ube.fr}}%
\thanks{$^{2}$MIS laboratory, University of Picardie Jules Verne, Amiens, France\newline
{\tt\small  guillaume.caron@u-picardie.fr}}%
\thanks{$^{3}$CNRS-AIST JRL, AIST, Tsukuba, Japan}%
\thanks{$^{4}$Artificial Intelligence Research Center, AIST, Tsukuba, Japan\newline
{\tt\small nevrez.imamoglu@aist.go.jp}}%
}

\maketitle
\thispagestyle{empty}
\pagestyle{empty}
\begin{abstract}

Saliency-guided coding allocates bits to the regions viewers are most likely
to attend, but accurate 360-degree saliency prediction is computationally
expensive, an obstacle for low-latency immersive streaming applications such as robot teleoperation.
We quantify how much saliency-prediction accuracy actually translates into compression gain.
Building on a saliency-guided rate--distortion-optimized (RDO) learned codec,
modified so that the training-time latent mask is driven by ground-truth saliency,
which yields a single, predictor-agnostic codec,
we evaluate saliency predictors spanning two orders of magnitude in compute on both the learned codec and H.265/HEVC.
The results show that higher saliency-prediction accuracy yields diminishing compression gains:
a lightweight predictor fast enough for live use (52 frames per second) captures $65\%$ of
the ground-truth oracle's bit-rate saving, a $3.3\times$ heavier one reaches $74\%$,
and a further $38\times$ increase in compute adds no measurable benefit, on either codec.
Therefore, lightweight predictors might also be a practical choice for latency-constrained 360-degree streaming.

\end{abstract}

\acresetall

\section{Introduction}
\label{sec:intro}

A growing range of applications must stream 360-degree video under limited bandwidth and latency constraints:
live video surveillance and remote site inspection, virtual attendance of live events, and, most demandingly, immersive teleoperation of robots.
Teleoperation is a particularly demanding application because it simultaneously stresses bandwidth, latency, and compute constraints; hence, we take teleoperation as the primary example.
Teleoperation keeps a human in the loop when robots cannot yet operate autonomously, such as in telesurgery, GPS-denied drone piloting, underwater and planetary exploration, and industrial maintenance. In these settings, the operator's situational awareness is often the limiting factor.
A conventional 2D camera
feed is frequently insufficient; 360-degree video viewed through a \ac{HMD} provides a more natural telepresence interface.
These three constraints are coupled: any pre- or post-processing introduced to save bandwidth itself adds to the end-to-end latency-compute budget.

\begin{figure}[!t]
\centering
\begin{tikzpicture}[font=\footnotesize, >={Stealth[length=1.8mm,width=1.5mm]},
  box/.style={draw=#1!70!black, fill=#1, rounded corners=1.5pt, align=center,
              inner sep=2.5pt, minimum height=8.5mm, line width=0.5pt},
  img/.style={inner sep=0pt, draw=black!55, line width=0.4pt},
  lab/.style={font=\scriptsize, text=black!75, align=center, inner sep=1pt},
  arr/.style={->, line width=0.6pt, black!70}]
  \def\tw{26mm}
  \node[img] (erp) at (0,0) {\includegraphics[width=\tw]{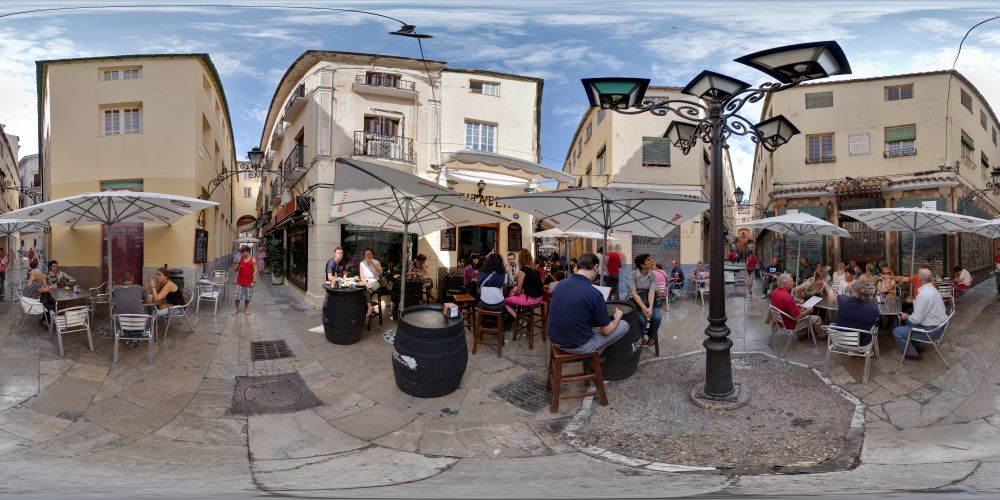}};
  \node[img] (sal) at (60mm,0) {\includegraphics[width=\tw]{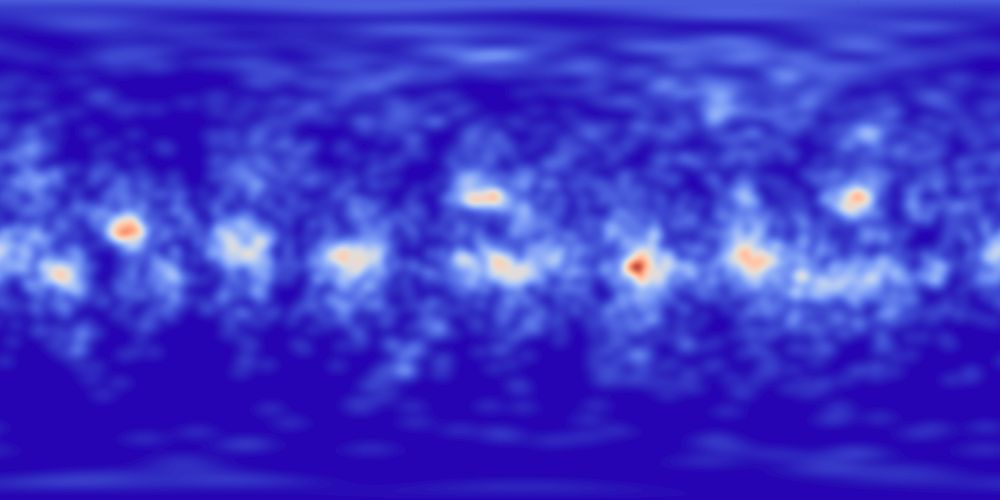}};
  \node[box=salb, text width=23mm] (pred) at (30mm,0) {Saliency\\predictor};
  \node[lab, above=0.6mm of erp] {360-degree image};
  \node[lab, above=0.6mm of sal] {saliency map};
  \node[lab, below=0.8mm of pred] {light $\leftrightarrow$ heavy\\less $\leftrightarrow$ more accurate};
  \draw[arr] (erp) -- (pred);
  \draw[arr] (pred) -- (sal);
  \node[box=elicg, text width=30mm] (enc) at (30mm,-19mm) {Saliency-guided\\encoder};
  \draw[arr] (erp.south) |- (enc.west);
  \draw[arr] (sal.south) |- (enc.east);
  \node[img] (bits) at (30mm,-36mm) {\includegraphics[width=34mm]{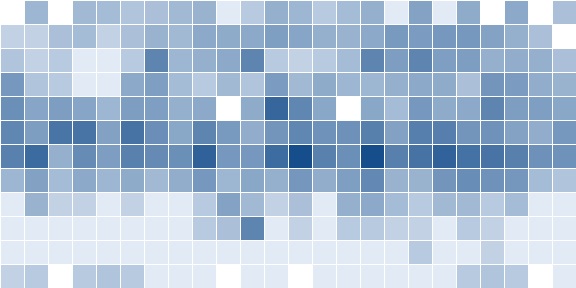}};
  \draw[arr] (enc) -- (bits);
  \node[lab, anchor=west, text width=22mm, align=left] at ([xshift=2mm]bits.east)
      {bit allocation:\\more bits where viewers look,\\fewer elsewhere};
  \node[lab, anchor=east, text width=20mm, align=right, text=black] at ([xshift=-2mm]bits.west)
      {\textbf{How accurate must the map be?}};
\end{tikzpicture}
\caption{Saliency-guided compression: a predicted saliency map steers the
encoder's bit allocation toward the regions viewers attend. This paper
measures how the accuracy and cost of the predictor translate into coding
gain.}
\label{fig:teaser}
\end{figure}

Three families of methods address 360-degree streaming: (i) Tiled/viewport streaming~\cite{viewport_saadallah} requires head-pose prediction and degrades under fast head motion;
(ii) foveated streaming~\cite{FOVeted} requires fast eye tracking and is purely reactive;
and (iii) saliency/content-aware~\cite{b_chiang} coding.
Saliency-aware coding is predictive and client(operator)-agnostic, but its cost moves to the saliency prediction model.
Both alternatives (viewport and foveated streaming) also allocate quality by viewing geometry alone, yet human attention is content-dependent~\cite{RUSSELL20141_attention_obj}.
Salient details remain perceivable in the periphery even when not gazed
upon, and purely gaze-driven quality reduction produces noticeable peripheral
artifacts~\cite{gao2025foveated}.
We therefore explored saliency-guided coding, which allocates bits by content (\cref{fig:teaser}) and degrades gracefully toward uniform-quality coding when the prediction is poor.
No 360-degree saliency predictor today is simultaneously accurate, real-time, and compression-friendly: the most accurate models rely on heavy computation on full equirectangular frames, and expensive temporal modelling, so predictors
that win on accuracy lose on latency and compute and do not fit a teleoperation workload.
This motivates our research question: how much does saliency-prediction accuracy actually translate into compression gain?

We conducted a controlled study to examine how accurate a saliency
model must be to meaningfully improve 360-degree image compression.
Our key findings are as follows:
\begin{itemize}
    \item
    We tested four different saliency sources on a fixed neural codec, ranging
    from a lightweight model ($\approx$305 GFLOPs) to a very heavy one
    ($\approx$39{,}000 GFLOPs), with correlation coefficients (CC) ranging from
    0.66 (moderate accuracy) to 1.0 (Oracle/perfect). The results show that a
    moderately accurate predictor already achieves about $70$--$75\%$ of the
    bitrate savings that would be possible with a perfect ``oracle'' (which
    uses ground-truth saliency). In contrast, a predictor that is 38 times
    more expensive and more accurate adds no measurable additional gain;
    \item The same ordering of predictors holds on a transform/block-based hybrid codec (HEVC saliency-RDO):
    there too the heaviest predictor adds nothing.
    A synthetic accuracy sweep further shows that current predictors sit below the accuracy at which the learned codec's gain levels off (\ac{CC}${}\approx{}0.8$--$0.9$).
\end{itemize}
The remainder of this paper is organized as follows: \cref{sec:bgrw}
reviews background and related work, \cref{sec:method} the methodology,
\cref{sec:setup} the experimental setup, \cref{sec:results} the results,
and \cref{sec:conclusion} concludes.

\section{Background and Related Work}
\label{sec:bgrw}

This section first defines the two core notions the paper rests on, image
coding and visual saliency, then covers 360-degree streaming and
saliency-aware coding (\cref{sec:rw:streaming}), 360-degree saliency
prediction (\cref{sec:rw:saliency}), saliency-guided compression in both
codec families (\cref{sec:rw:salcoding}), and the gap we address
(\cref{sec:rw:gap}).
An \textbf{Image codec} compresses a picture into a bit-stream by transforming,
quantizing, and entropy-coding its content, trading rate (the bit-rate, i.e.,
the number of coded bits per pixel) against distortion.
Transform/block-based hybrid codecs, built from hand-designed transforms, such as JPEG and the \textit{intra} mode of H.265/HEVC, set per-block quality with a \ac{QP},
whereas learned codecs train an auto-encoder end-to-end under a rate-distortion
objective~\cite{b_balle,b_elic}, with variable-rate variants avoiding one model per rate point~\cite{b_qvrf}.
A \textbf{Saliency model}, in turn, predicts where human attention falls, producing a per-pixel fixation-probability map whose accuracy is scored against eye-tracking ground truth using the \ac{CC}, \ac{NSS}, \ac{KLD}, and \ac{AUCJ}.
Both problems are harder for \acp{ODI}: the \ac{ERP} format
oversamples the poles and viewing is equator-biased, so quality is assessed
with sphere-aware metrics such as \ac{WSPSNR}~\cite{b_wspsnr}, and the most
accurate 360-degree saliency predictors are also the most computationally
expensive.

\subsection{360-degree Streaming and Saliency-Aware Coding}
\label{sec:rw:streaming}
Of the three streaming families introduced in \cref{sec:intro}, we focus on
saliency and content-aware coding because its latency-critical cost is upstream of transmission and tunable through model choice, which suits a teleoperation workload:
tiled/viewport and foveated streaming incur additional latency by tracking the operator's head or gaze,
whereas saliency-aware (predictive) coding biases the encoder's bit allocation
toward regions attention is likely to visit before the stream leaves the robot.
Saliency-aware coding has been used inside both block-based hybrid and learned codecs: in HEVC,
Hadizadeh and Baji\'c~\cite{b_hadizadeh} and Chiang \textit{et al.}~\cite{b_chiang} adjust the per-block \ac{QP} from a saliency map;
in learned codecs, the saliency map modulates the distortion loss and (or) is used as
a latent-space mask~\cite{b_gungordu,roi_salmap_blur,b_patel}.

\subsection{Saliency Prediction for 360-degree Content}
\label{sec:rw:saliency}
In recent years, a number of 360-degree saliency predictors have been proposed, with different accuracy and compute footprints.
SalGAN360~\cite{b_salgan360} extends a 2D \ac{GAN}-based predictor to
omnidirectional content by predicting on cube-map faces and fusing the projections.
ATSal~\cite{b_atsal} uses a dual-stream design that combines a global attention stream with a local recurrent stream on cube-map patches
(only the ``attention'' stream of ATSal is used as a predictor in our experiments; see \cref{sec:method:sources}).
The dual-stream detector of G\"ung\"ord\"u and Tekalp~\cite{b_gungordu} essentially extends ATSal by coupling Attention and Expert streams with coordinate-refinement~\cite{_CRN} and fusion sub-networks.
We employ this detector to generate saliency maps for the baseline learned codec adopted in this study.

\subsection{Saliency-Guided Compression}
\label{sec:rw:salcoding}
Saliency-guided compression has been studied in both codec families.
In learned compression, \ac{RDO} beyond the training objective is a topic in
its own right: Sch\"afer \textit{et al.}~\cite{_RDO_optim} optimize the
quantized latents of a trained network per image at encoding time, and Brand
\textit{et al.}~\cite{_Tru_RDO} obtain a block-partitioning-like dynamic
\ac{RDO} by training a hierarchical autoencoder against a transmitted
latent-space mask. Saliency-guided codecs instead make the objective itself
attention-aware: learned variants typically combine a saliency-weighted
distortion loss with latent-space masking, so the rate term itself becomes
saliency-aware.
Ma \textit{et al.}~\cite{roi_salmap_blur} introduced an \ac{ROI} variant of end-to-end learned image compression;
Patel \textit{et al.}~\cite{b_patel} applied saliency-driven perceptual coding to 2D images.
On the block-based hybrid codec side,
Chiang \textit{et al.}~\cite{b_chiang} implement a per-\ac{CTU} $\Delta$QP map for 360-degree HEVC.
Interestingly, Cai \textit{et al.}~\cite{b_cai} observed qualitatively that in 2D ROI image coding,
``more accurate'' region prediction does not monotonically translate into coding gain.
Closest to the present study, for 360-degree content G\"ung\"ord\"u and Tekalp~\cite{b_gungordu} combine \ac{ELIC} with a \ac{QVRF} variable-rate mechanism and a dual-stream saliency predictor,
the configuration we build on (\cref{sec:method}).
They compare several 360-degree saliency predictors on prediction accuracy alone and use a single saliency prediction model in compression,
leaving the effect of saliency prediction accuracy on coding gain unexamined.

\subsection{Research Gap}
\label{sec:rw:gap}
Although saliency and \ac{ROI}-guided coding has been applied to
360-degree compression in both the learned and block-based hybrid codec
families, no existing work examines the added cost of saliency prediction in
the 360-degree domain under low-latency constraints.
As just noted for the closest baseline~\cite{b_gungordu},
none relates predictor accuracy or compute to the resulting \ac{RDO} coding gain.
The open question is therefore quantitative:
how much coding gain does saliency accuracy actually yield for 360-degree compression
across a wide accuracy--complexity range, and is that relationship codec-specific?
We answer it by comparing four saliency sources, and a synthetic accuracy
sweep, on a fixed predictor-agnostic learned codec and on HEVC saliency-RDO.
To our knowledge this is the first study to quantify this relationship for
360-degree content, and the answer is not the expected monotone one: the
most accurate predictor is not the best.

\section{Methodology}
\label{sec:method}
Addressing this gap requires a controlled experiment in which saliency sources of varying accuracy and computational cost are evaluated on fixed codecs, ensuring that differences in coding gain can be attributed solely to the saliency maps.
\Cref{sec:method:overview} outlines the pipeline, \cref{sec:method:sources}
the saliency sources, \cref{sec:method:learned,sec:method:train} the learned
codec and the training protocol that makes it predictor-agnostic, and
\cref{sec:method:hevc} the HEVC configuration.

\begin{figure*}[!htp]
\centering
\includegraphics[width=\textwidth]{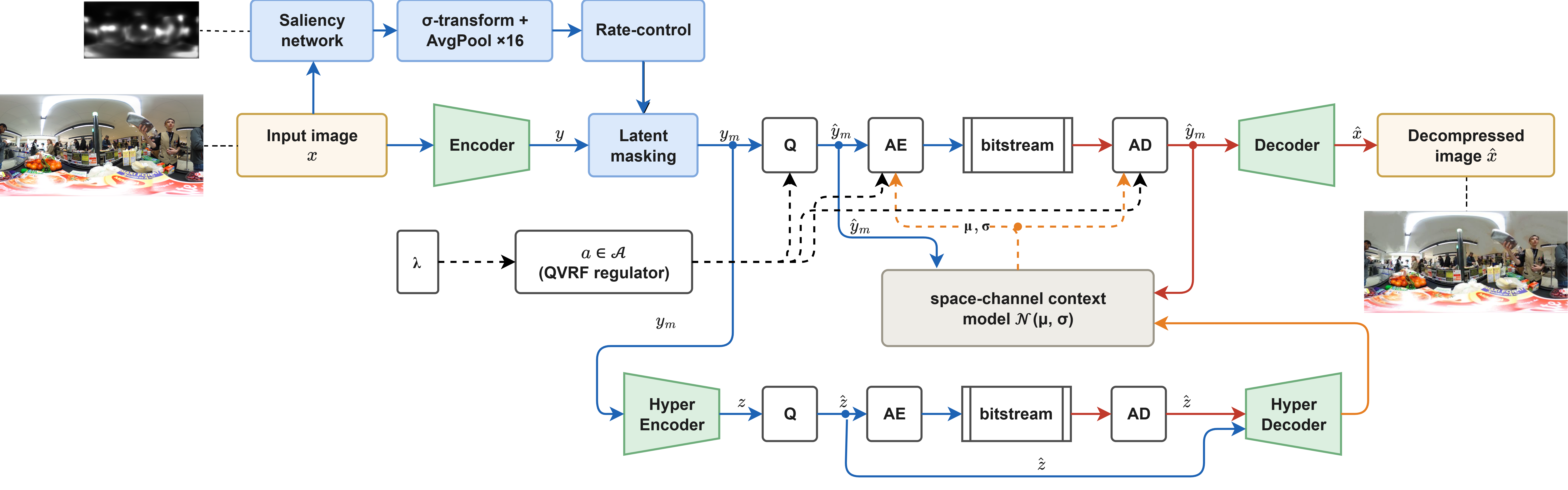}
\caption{Saliency-guided learned codec used in this study (ELIC-QVRF variable-rate control and latent-space saliency masking), following the
design of~\cite{b_gungordu}.
The saliency map is $\sigma$-transformed, average-pooled ($\times 16$) to the latent resolution,
and passed through rate control before masking the latent $y$.
The space-channel context model predicts the mean and scale $(\mu,\sigma)$ of the
conditional Gaussian entropy model used by the arithmetic encoder/decoder (AE/AD).}
\label{fig:arch}
\end{figure*}

\subsection{System Overview}
\label{sec:method:overview}
The system operates as a linear pipeline: a saliency predictor computes a
per-pixel attention map from the captured \ac{ERP} frame, a saliency-guided
encoder uses that map to bias its rate allocation, and the bit-stream is
transmitted, decoded, and rendered in the operator's \ac{HMD}.
The compression stage itself (\cref{fig:arch}) couples two networks across a
single interface, the saliency map. A saliency network estimates where
attention falls, and the learned codec consumes the map twice: as a weight on
the distortion term of its training objective and as a multiplicative mask on
part of the latent representation, so that bits are steered toward salient
regions. The map is used only at the encoder and is never transmitted, leaving
the decoder unchanged~\cite{b_gungordu}. The same map can instead drive a per-block \ac{QP} offset in a standard
HEVC encoder (the block-based hybrid codec configuration of the study).

\subsection{Saliency Predictors}
\label{sec:method:sources}
We evaluate three saliency predictors of increasing accuracy and cost, the
attention stream of ATSal~\cite{b_atsal} (its attention-only output, not the
full dual-stream model), the reproduced dual-stream detector
of~\cite{b_gungordu}, and the heavier SalGAN360~\cite{b_salgan360}, plus a \ac{GT} oracle as the reference.
Their quality and compute are reported alongside the coding results in \cref{tab:saliency};
these sources span the accuracy--compute axis along which we measure coding gain.
Throughput (FPS, \cref{tab:saliency}) is measured on an NVIDIA RTX A4500
at each predictor's working resolution ($640\times320$ for ATSal and the
dual-stream model, $2048\times1024$ for SalGAN360).
Throughout, we call a predictor \emph{real-time} when its throughput exceeds the 30-FPS capture rate of typical 360-degree rigs:
by this criterion only the attention-only ATSal model (52 FPS) qualifies,
while the dual-stream model (4.8 FPS) is near-real-time and SalGAN360 (0.04 FPS) is unusable for live streaming.

\subsection{Learned Codec: ELIC--QVRF with Saliency Masking}
\label{sec:method:learned}
We build our compression pipeline on the open-source ELIC--QVRF
implementation~\cite{b_elicqvrf}, coupling the ELIC backbone~\cite{b_elic}
with QVRF variable-rate control~\cite{b_qvrf}, and add saliency-guided
\ac{RDO} and latent masking
following~\cite{b_gungordu,roi_salmap_blur} (illustrated in \cref{fig:arch}).
Here the encoder--decoder pair is ELIC's learned transform between image
$x$ and latent $y$; the hyper-encoder and hyper-decoder
transmit a small side-latent $z$ capturing the spatial structure of
$y$~\cite{b_balle}; and the space-channel context model combines
the decoded $\hat{z}$ with already-decoded latent groups to predict the
entropy-model parameters $(\mu,\sigma)$, whose internals follow
ELIC~\cite{b_elic}. The codec is trained
with a saliency-weighted \ac{MSE} distortion, and bits are steered toward salient
regions by scaling the non-preserved latent channels with a
$\times16$-downsampled saliency mask (channels $0$--$47$ preserved, $48$--$191$
saliency-weighted).
The codec offers a discrete set of rate--quality operating points, indexed
by $s$ and obtained by scaling the latent with the learned QVRF gain vector,
which we call the \emph{rate ladder}.

During training we drive this latent mask with the \ac{GT} saliency map,
\begin{equation}\label{eq:gt_sat}
\mathrm{Mask} =
\frac{\mathrm{AvgPool}_{16}\big(\sigma_{\text{shifted}}(S_{\mathrm{GT}})\big)
+\alpha}{\alpha},
\quad \alpha = 1,
\end{equation}
where $S_{\mathrm{GT}}$ is the \ac{GT} saliency map, and substitute a
predictor only at inference; the rationale and its consequences
are detailed in \cref{sec:method:train}. Here
$\sigma_{\text{shifted}}(t)=\sigma(10\,t-2.5)$ is a rescaled sigmoid: applied
literally, the sigmoid mask of~\cite{b_gungordu} saturates on our data
($50$--$75\%$ of pixels above $0.95$), and the rescaling restores the contrast
between salient and non-salient regions. The residual term $(\cdot+\alpha)/\alpha$
keeps the mask in $[1,2]$, so the scaled latent channels are amplified in
salient regions while non-salient regions remain near unity.

The training objective is the rate-distortion Lagrangian $L = D + \lambda R$
standard in end-to-end learned compression~\cite{b_balle}, with the distortion
term made saliency-aware following~\cite{b_gungordu}:
\begin{equation}\label{eq:loss}
L = \lambda\, L_{\text{Sal-MSE}} + L_{\text{bpp}},\quad
L_{\text{Sal-MSE}} = \frac{\sum_{i=1}^{N} S_i\,(x_i-\hat{x}_i)^2}{\sum_{i} S_i},
\end{equation}
where $x$ and $\hat{x}$ are the original and reconstructed images, $S$ is the
\ac{GT} saliency map passed through the same sigmoid transform as the latent
mask (\cref{eq:gt_sat}), $N$ is the number of pixels, and
$L_{\text{bpp}}$ is the estimated rate of the latent
representation~\cite{b_gungordu}. The three-stage training schedule is
detailed in \cref{sec:setup:training}.

\subsection{Training the Latent Mask with Ground-Truth Saliency}
\label{sec:method:train}
The baseline and the \ac{ROI} codecs we build upon~\cite{b_gungordu,roi_salmap_blur} hold
the saliency (or \ac{ROI}) network fixed/essentially-frozen while the compression network is being trained,
and feed its \emph{predicted} saliency map into the training-time latent masking
(\cite[Eq.~1]{b_gungordu}). Although the two networks train separately,
the training-time mask couples the codec to the predictor's error statistics (\cite[Eqs.~1--3]{b_gungordu}),
so changing the predictor requires retraining the codec.
We instead replace the \textit{Saliency Network} (\cref{fig:arch}) with \ac{GT} saliency masks during training time (\cref{eq:gt_sat}) and
substitute a Saliency predictor network only at inference.
The learned latent masking is then independent of any predictor,
so one \ac{GT}-trained compression checkpoint serves every saliency source.
This makes the controlled comparison at the heart of this study practical:
every predictor is scored on the same codec, not affected by per-predictor training.

The choice is also marginally better for coding. Against a predictor-in-loop
variant evaluated with the same inference mask, the \ac{GT}-trained codec is
consistently ahead on paired per-image SAL-PSNR BD-rate (median $-1.5\%$,
$20/25$ images, sign-test $p\approx0.004$), a small but statistically reliable
margin.
We attribute this to \ac{GT} saliency providing clean, noise-free
signal to the latent-mask computation during training. The masks presented
during training are known to shape the trained codec. Brand \textit{et
al.}~\cite{_Tru_RDO}, for instance, replaced the random training-time masks
of their hierarchical codec with content-derived masks that approximate the
inference-time \ac{RDO} outcome, and reported large BD-rate gains from that
change alone. Our protocol applies the same principle with the cleanest mask
available, the \ac{GT} saliency map.

\subsection{Block-Based Hybrid Codec: HEVC Saliency-RDO}
\label{sec:method:hevc}
To evaluate whether the conclusion is codec-specific, we replicate the
experiment in HEVC (kvazaar~\cite{b_kvazaar}, all-intra) with a per-\ac{CTU}
$\Delta$QP map derived from saliency, following Chiang \textit{et
al.}~\cite{b_chiang}. Each \ac{CTU} receives a weight $w = w_c\, w_s$, where
$w_c$ is the content weight (the mean of the saliency map over the \ac{CTU})
and $w_s$ the latitude-dependent spherical weight of the \ac{ERP} format; the
\ac{QP} offset is then $\Delta\mathrm{QP} = -3\log_2(w/\bar{w})$, with
$\bar{w}$ the frame-mean weight, clamped to $\pm3$, against a flat-\ac{QP}
HEVC anchor.

\section{Experimental Setup}
\label{sec:setup}
We follow the same experimental methodology as the
baseline~\cite{b_gungordu}: the Salient360! ICME-2017 test split (25
omnidirectional images, ODIs) at native equirectangular projection resolution
of $2048\times1024$ pixels, using the Salient360! 2018 ground-truth (GT)
maps, with BD-rate and BD-PSNR integrated over rate-ladder indices
$s=0\ldots5$ (the ladder of \cref{sec:method:learned}).
Because the original source code is unavailable and the 2017 saliency set was
retired and replaced by a non-identical 2018 release with different \ac{GT}, we reproduce the baseline
at the level of \emph{trend} rather than exact figures: the SAL-PSNR trend
reproduces, with residual differences we attribute to this dataset shift,
whereas the baseline's \ac{WSPSNR} gain does not (\cref{sec:results:allocation}).
Evaluations ran on an NVIDIA RTX A4500 (20\,GB).
\Cref{sec:setup:training} details the training schedule;
\cref{sec:method:metrics} defines the evaluation metrics.

\subsection{Training Details}
\label{sec:setup:training}
Rate control follows the three-stage QVRF schedule~\cite{b_qvrf,b_gungordu}:
the network is first optimized at a single fixed rate--distortion trade-off
$\lambda=0.18$, the \emph{anchor}, i.e., the highest-rate operating point of
the ladder, trained first so the later stages start from a strong model; the
quantization regulator vector is then refined jointly over the eight-point set
$\lambda\in\{0.0018,0.0035,0.0067,0.013,0.025,0.0483,0.0932,0.18\}$
inherited from the QVRF schedule~\cite{b_qvrf},
defining rate indices $s=0\ldots7$; and a final stage fine-tunes the network
with straight-through estimation, applied to the image latent but not to the
hyperprior latent: exempting the hyperprior was necessary for numerically
stable training and avoids a further $\sim$$0.1$ dB loss in spherical BD-PSNR.
Mirroring the baseline schedule~\cite{b_gungordu}, each stage runs a 2D
pre-training phase on the SALICON training split~\cite{b_salicon}
($512\times512$ pixels, $200$k iterations, batch $16$, Adam with learning rate
$10^{-4}$) followed by 360-degree fine-tuning on the 60-image Salient360!
training split~\cite{b_salient360_images}
(\ac{ERP} $2048\times1024$ with
random $1024\times512$ or $512\times1024$ crops and horizontal flips, $10$k
iterations, batch $8$, learning rate $10^{-5}$). Each stage takes roughly one
to one and a half days on a single GPU (NVIDIA V100S or A100).

\subsection{Evaluation Metrics}
\label{sec:method:metrics}
We report compression quality primarily with SAL-PSNR~\cite{b_salpsnr},
 a saliency-weighted metric that emphasises regions where human attention is likely to focus,
 and, as a spherical-aware companion, \ac{WSSALPSNR}, which accounts for the distortion introduced by the equirectangular projection format.
 Coding efficiency is summarized by \ac{BDRATE} and
\acs{BDPSNR}~\cite{b_bjontegaard} integrated over rate-ladder indices
$s=0\ldots5$; a negative BD-rate denotes a bit-rate saving at equal quality.
SAL-PSNR uses the \ac{GT} map as its weight, so the \ac{GT} oracle is by
construction the reference this metric favors.
WS-PSNR~\cite{b_wspsnr}, which is independent of the \ac{GT} map, is
reported alongside (\cref{tab:bd_main}).
Saliency-prediction accuracy is measured with \ac{CC}, \ac{KLD},
\ac{NSS}, and \ac{AUCJ} under a fixed protocol (native \ac{GT} resolution; spherical
weighting on \ac{CC} and \ac{KLD}) using the toolbox provided with the dataset~\cite{b_salient360_toolbox, b_salient360_images}.

\section{Experimental Results}
\label{sec:results}
This section answers our central question, how much coding gain saliency
accuracy actually yields for 360-degree compression and whether the
relationship is codec-specific: \cref{sec:results:diminishing} presents the
accuracy-versus-gain comparison on the learned codec,
\cref{sec:results:hevc} validates the trend on HEVC,
\cref{sec:results:crosscodec} compares both codecs against a common anchor,
\cref{sec:results:allocation} analyses the bit allocation,
\cref{sec:results:qual} gives a visual comparison, and
\cref{sec:results:sweep} extends the accuracy axis synthetically.

\subsection{Saliency Accuracy vs.\ Coding Gain}
\label{sec:results:diminishing}
\begin{figure}[t]
\centering
\includegraphics[width=\columnwidth]{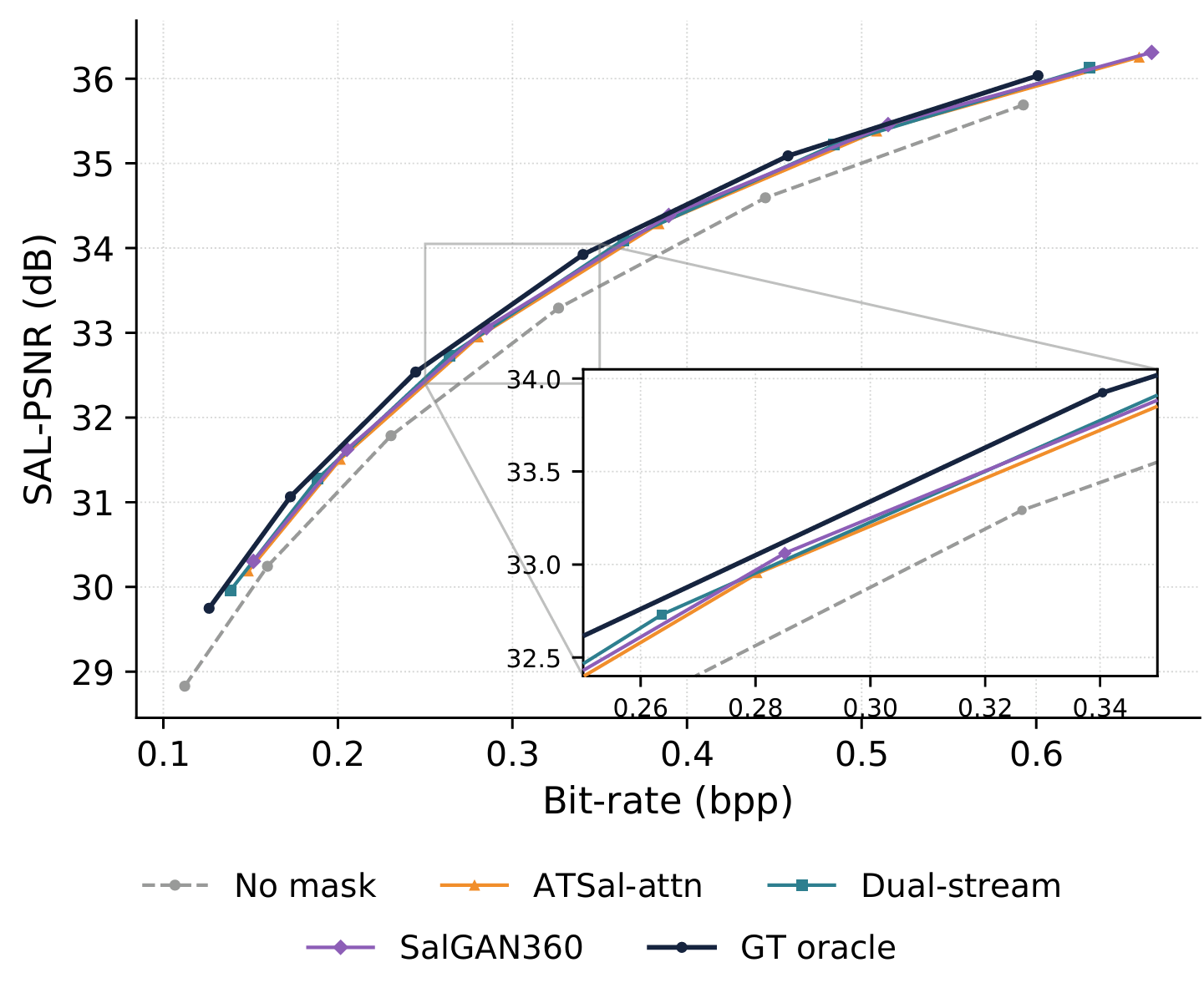}
\caption{Mean SAL-PSNR rate-distortion over the 25 test ODIs on the learned codec,
by saliency source (inset: zoom on $0.25$--$0.35$ bpp).}
\label{fig:rd}
\end{figure}

\begin{figure}[t]
\centering
\includegraphics[width=0.96\columnwidth]{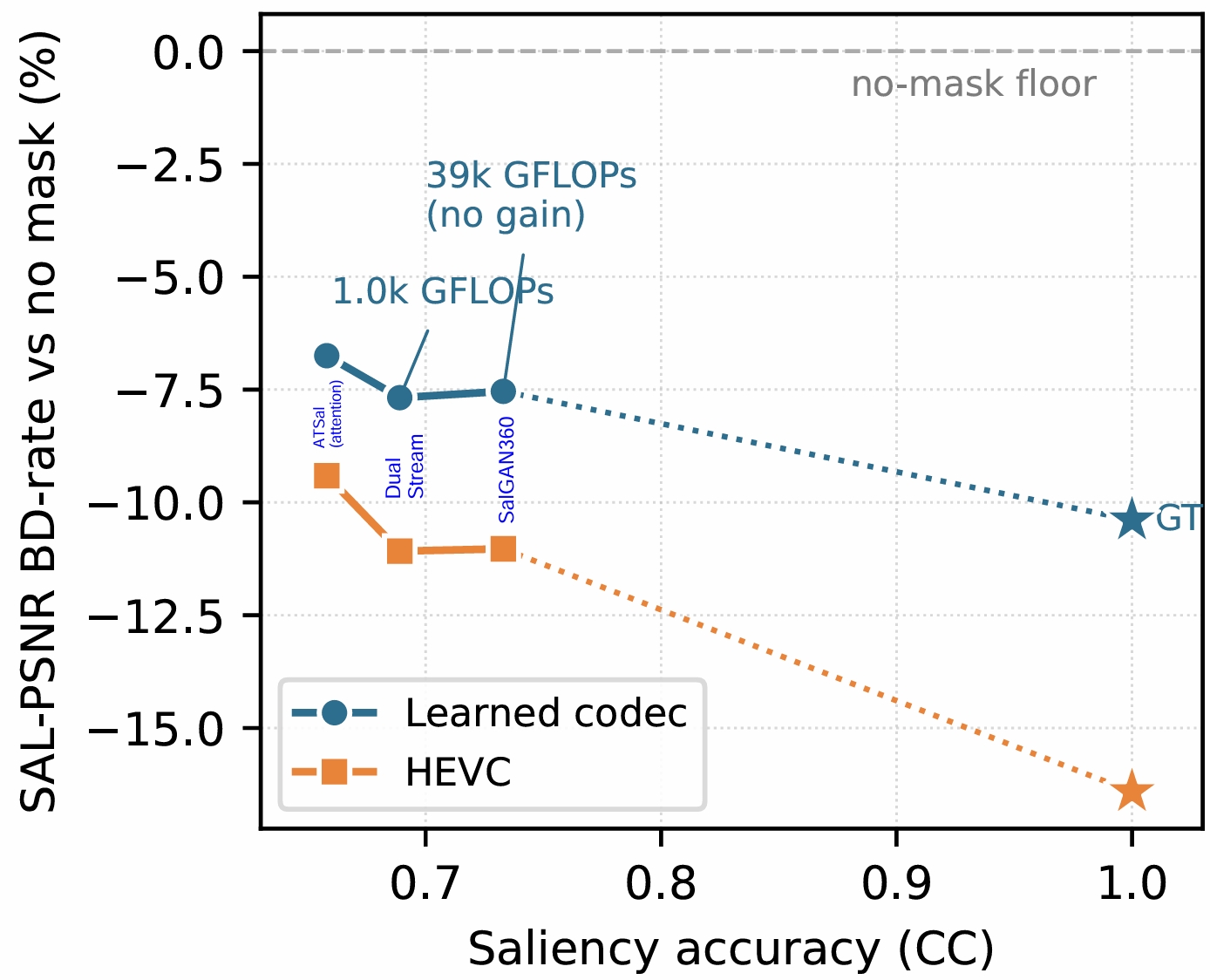}
\caption{Mean SAL-PSNR BD-rate versus predictor accuracy on both codecs.}
\label{fig:diminishing}
{\par\smallskip\footnotesize Note: each codec is measured against its own no-saliency reference.\par}
\end{figure}

All predictor masks are evaluated on the same \ac{GT}-trained codec, with the \ac{GT}
oracle as the reference (\cref{fig:rd}, \cref{tab:bd_main}).
Every saliency predictor sits between the no-mask baseline and the oracle, but
in a narrow band: the predictors are nearly indistinguishable on the
rate-distortion plane (\cref{fig:rd}; the \ac{WSSALPSNR} curves show the
same ordering and are omitted for space). Coding gain rises with saliency
accuracy yet saturates well below the oracle: the mid-accuracy dual-stream
predictor reaches a SAL-PSNR BD-rate of $-7.68\%$, about $74\%$ of the oracle's
$-10.39\%$. The decisive observation is that SalGAN360, the highest-\ac{CC} source at roughly $38\times$ the compute of the dual-stream model, does \emph{not} improve
on it ($-7.54\%$; \cref{fig:diminishing}), while the attention-only model
trails ($-6.75\%$). Beyond a mid-accuracy threshold, additional saliency
accuracy and compute yield no measurable coding gain.

Two readings of saliency accuracy are worth separating. \emph{Across}
predictors, the mean gain saturates and even inverts the \ac{CC} order, as above.
This does not mean accuracy is irrelevant: \emph{within} a fixed predictor,
across the 25 test \acp{ODI}, images on which saliency is predicted more accurately
receive larger savings. Correlating each image's SAL-PSNR saving (dual-stream
mask vs.\ no mask) against that image's saliency accuracy gives a positive,
significant association for every accuracy metric: \ac{CC} (Spearman $\rho=0.48$,
$p=0.01$), with \ac{NSS} ($\rho=0.57$) and \ac{AUCJ} ($\rho=0.61$) somewhat stronger, and
\ac{KLD} weakest and not significant. Saliency accuracy is thus broadly \emph{indicative} of per-image coding
gain, with no single metric standing out as a decisive predictor.

\begin{table*}[!t]
\caption{Saliency sources: prediction accuracy, compute cost, and BD-rate
on both codecs. Lower BD-rate is better.}
\label{tab:saliency}\label{tab:bd_main}
\centering
\small\setlength{\tabcolsep}{4.2pt}
\begin{tabular}{lcccccccccc}
\toprule
 & & & & & & & \multicolumn{3}{c}{vs own ref.$\downarrow$} & vs anchor$\downarrow$ \\
\cmidrule(lr){8-10}\cmidrule(lr){11-11}
 & & & & & & & \multicolumn{2}{c}{SAL-PSNR} & WS-PSNR & SAL-PSNR \\
Saliency Mask Source & CC$\uparrow$ & KLD$\downarrow$ & NSS$\uparrow$ & AUC-J$\uparrow$ & GFLOPs$\downarrow$ & FPS$\uparrow$ & Learned & HEVC & Learned & Learned \\
\midrule
GT (oracle)            & 1.000 & 0.000 & --    & 1.000 & --       & --   & $-10.39\%$ & $-16.41\%$ & $+2.74\%$ & $-53.9\%$ \\
SalGAN360              & 0.733 & 0.369 & 1.310 & 0.798 & 38\,912  & 0.04 & $-7.54\%$ & $-11.03\%$ & $+5.34\%$ & $-52.8\%$ \\
Dual-stream (reprod.) & 0.689 & 0.400 & 1.200 & 0.774 & 1\,012.7 & 4.8 & $-7.68\%$ & $-11.08\%$ & $+4.11\%$ & $-52.6\%$ \\
ATSal (attention-only) & 0.658 & 0.428 & 1.161 & 0.772 & 304.8    & 52.4 & $-6.75\%$ & $-9.41\%$ & $+4.84\%$ & $-52.3\%$ \\
\midrule
ELIC-QVRF (no mask)    & --    & --    & --    & --    & --       & --   & $0\%$ & -- & $0\%$ & $-47.4\%$ \\
\bottomrule
\end{tabular}
{\par\smallskip\footnotesize Note: ``vs own ref.'' uses each codec's
no-saliency reference (learned: ELIC-QVRF~\cite{b_elicqvrf} no mask; HEVC: flat-QP anchor), comparable
within a codec only; FPS on an RTX A4500 (fp32, batch 1, working resolution); the last
column uses the common HEVC anchor.\par}
\end{table*}

\begin{figure*}[!b]
\centering
\includegraphics[width=\textwidth]{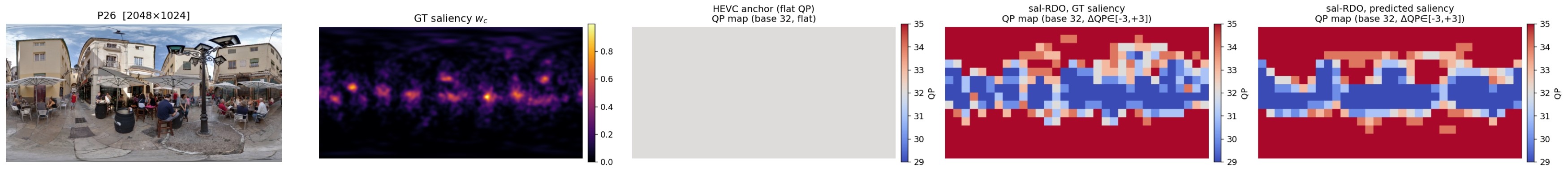}
\caption{Per-\ac{CTU} QP maps on P26 (base QP 32): flat-QP anchor vs.\
saliency-\ac{RDO} with \ac{GT} and predicted saliency
($\Delta$QP $\in[-3,+3]$).}
\label{fig:qpmaps}
{\par\smallskip\footnotesize Note: salient \acp{CTU} receive a lower QP (more
bits); the predicted-saliency allocation closely follows the \ac{GT} one.\par}
\end{figure*}

\subsection{Validation on a Block-Based Hybrid Codec}
\label{sec:results:hevc}
Under HEVC saliency-\ac{RDO} the pattern repeats: the oracle reaches $-16.41\%$ SAL-PSNR BD-rate,
the dual-stream and SalGAN360 predictors capture $\sim$$68\%$ of it and
are practically identical ($-11.08\%$ vs $-11.03\%$), and the
attention-only model trails. The cost in \ac{WSPSNR} is near-zero for the
predicted maps ($\pm0.4\%$ BD-rate; $+1.9\%$ for the oracle): saliency-\ac{RDO} trades pole and background quality for foreground at
almost no global-PSNR penalty, and the foreground receives more bits per
pixel: from kvazaar's per-\ac{CTU} bit log, the foreground-to-background
bpp ratio rises by $27$--$37\%$ over the flat-QP anchor with the \ac{GT}
oracle and by $20$--$31\%$ with the predictors, the effect being largest at
low rates (\cref{fig:qpmaps}).
This directionally matches~\cite{b_chiang}, though the numbers are not
commensurable (different content, saliency model, and saliency-PSNR
variant).

\begin{figure*}[!t]
\centering
\includegraphics[width=\textwidth]{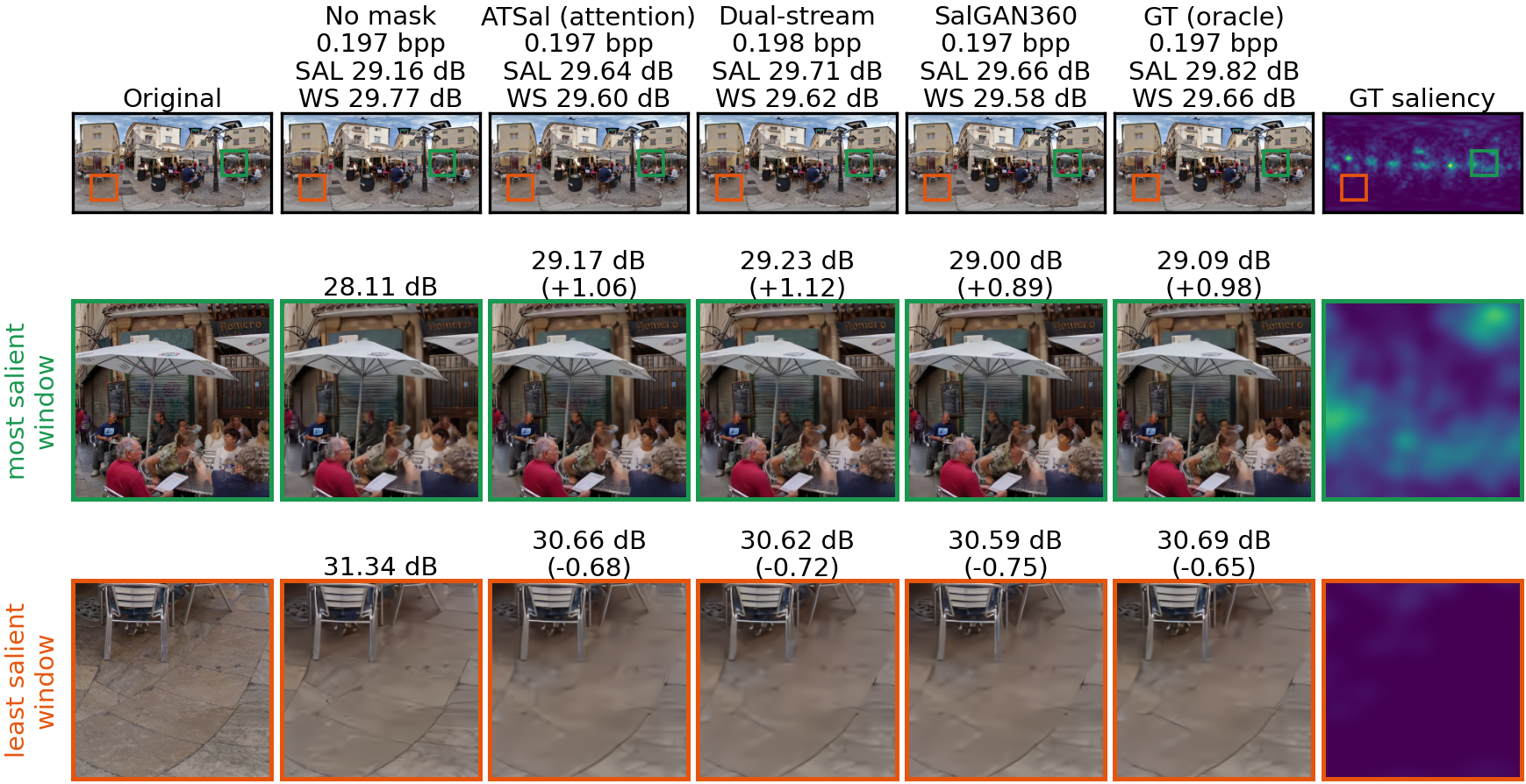}
\caption{Decoded P26 at matched bit-rate ($0.197$ bpp; masked codecs
rate-matched by gain interpolation), for each saliency source and the
\ac{GT} oracle. Column titles give whole-image bpp, SAL-PSNR and WS-PSNR.
Green/orange: the $256\times256$ window with the highest/lowest \ac{GT}
saliency, with crop PSNR and its change against no mask.}

\label{fig:qual}
\end{figure*}

\begin{figure}[!t]
\centering
\includegraphics[width=\columnwidth]{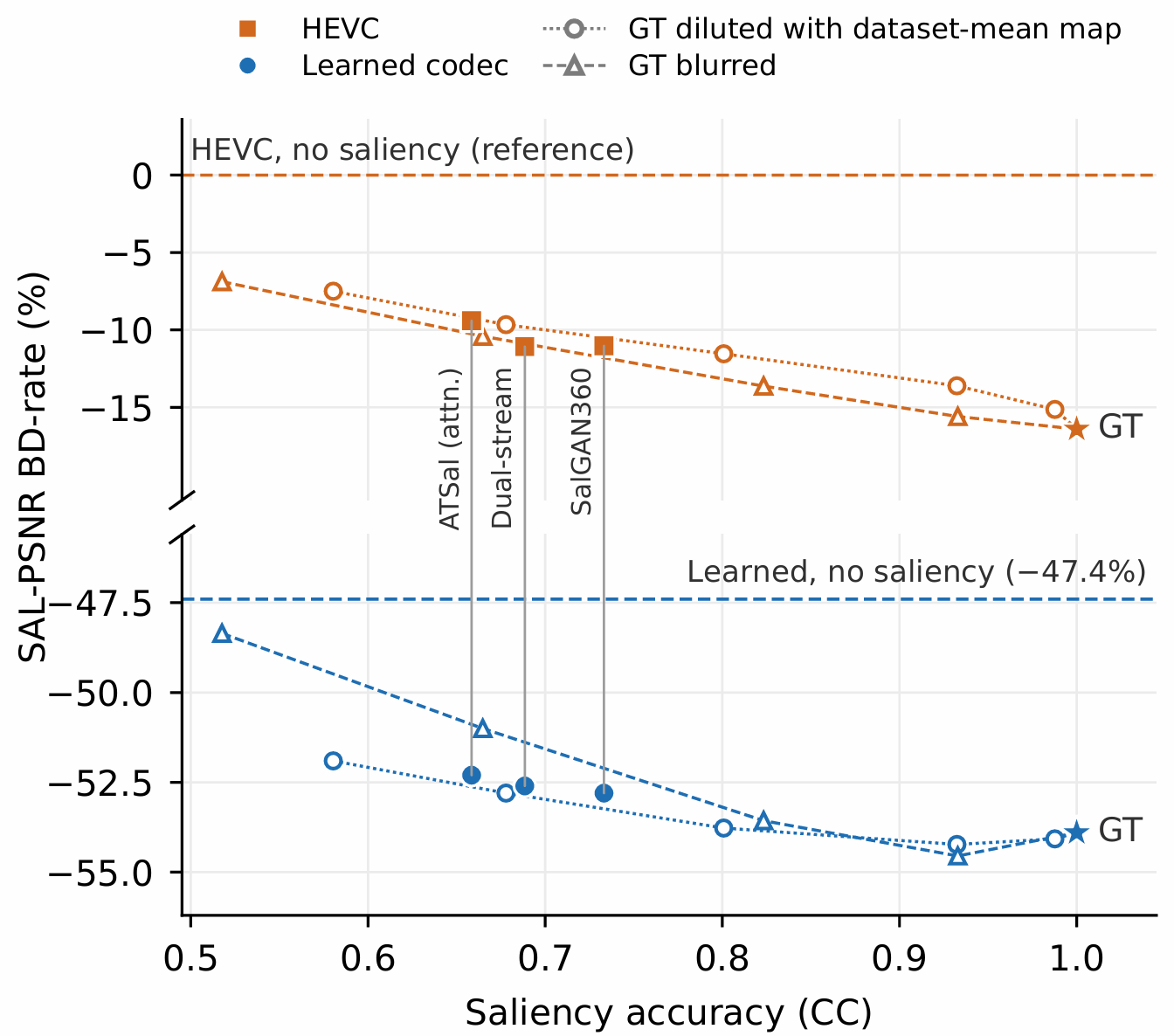}
\caption{SAL-PSNR BD-rate relative to flat-QP HEVC versus saliency accuracy
(lower is better; the two panels use different vertical scales):
synthetically degraded \ac{GT} maps (hollow) and real predictors (filled).}
\label{fig:sweep}
\end{figure}

\subsection{Cross-Codec Context}
\label{sec:results:crosscodec}
Against a common HEVC anchor (\cref{tab:bd_main}, last column), switching from HEVC to
the learned codec saves $47.4\%$ BD-rate before any saliency is applied
(\cref{tab:bd_main}, last row),
while saliency alone on HEVC saves $9$--$11\%$; applying both reaches $52$--$54\%$.
The codec switch alone is worth roughly three times what even oracle saliency
gains on HEVC ($-16.4\%$), underlining the potential of learned codecs for
bandwidth-limited 360-degree streaming, albeit at a higher compute cost (below).
While the benefits of these two compression techniques overlap;
a near-real-time predictor on the learned codec ($-52.6\%$) lands within
about a point of the learned oracle ($-53.9\%$).
On the same RTX A4500 at $2048\times1024$, the learned codec
encodes in $756$\,ms and decodes in $533$\,ms per frame (median over the 25
\acp{ODI}; mostly CPU entropy coding), and kvazaar all-intra encodes in
$135$\,ms on an Intel Xeon Silver 4210R. The compute path saliency${}+{}$encode${}+{}$decode is
thus ${\approx}1.3$\,s with ATSal and ${\approx}1.5$\,s with the
dual-stream model, dominated by the codec; network and rendering latency
are outside the scope of this image-coding study.

\subsection{Bit Allocation Analysis}
\label{sec:results:allocation}
Saliency masking redistributes bits toward salient regions at a small global
cost in \ac{WSPSNR}: $-0.1$ to $-0.2$\,dB BD-PSNR ($+2.7$ to $+5.3\%$ BD-rate,
\cref{tab:bd_main}) on the learned codec, and within $\pm0.4\%$ BD-rate on
HEVC for the predicted maps ($+1.9\%$ for the oracle).
Unlike G\"ung\"ord\"u and Tekalp~\cite{b_gungordu}, who report a $0.42$\,dB \ac{WSPSNR} gain, we observe a small loss,
in line with Cai \textit{et al.}~\cite{b_cai}, who note that \ac{ROI} coding trades whole-image quality for a high-quality \ac{ROI} at a given rate.
We could reproduce the baseline's reported joint WS$+$SAL
improvement~\cite{b_gungordu} only by blending the training-time mask with a
mild equatorial prior, and the effect did not survive the final
straight-through fine-tune ($+0.39\rightarrow-0.08$\,dB WS BD-PSNR); we
attribute the residual gap to fine-tuning and dataset shift (between Salient360!
2017 and Salient360! 2018 GT maps).

\subsection{Qualitative Results}
\label{sec:results:qual}
At a matched bit-rate of $0.197$\,bpp (\cref{fig:qual}), the three
predictors raise whole-image SAL-PSNR by $0.5$\,dB over the unmasked codec
($29.64$--$29.71$ against $29.16$\,dB) and the \ac{GT} oracle by
$0.66$\,dB ($29.82$\,dB), each at a $0.11$--$0.19$\,dB cost in WS-PSNR.
Locally the trade is larger: in the most salient window the masked codecs
gain $0.9$--$1.1$\,dB, while in the least salient window they lose
$0.65$--$0.75$\,dB, i.e., saliency guidance moves quality rather than
adding it. All saliency sources, the oracle included, are visually nearly
indistinguishable; in this window the oracle ($+0.98$\,dB) and SalGAN360
($+0.89$\,dB) gain no more than the lighter predictors, so beyond a
moderate level of saliency accuracy, better prediction brings no
observable gain in reconstruction quality.

\subsection{Synthetic Accuracy Sweep}
\label{sec:results:sweep}
To cover the accuracy axis beyond the three predictors (\ac{CC}
$0.66$--$0.73$), we degrade the \ac{GT} map synthetically,
blending it with the mean \ac{GT} map of the other test images, a
content-independent prior ($\beta\in\{0.25, 0.5, 0.75, 0.9, 1\}$), and
blurring it (Gaussian $\sigma\in\{32, 64, 128, 256\}$ px), which yields
maps with \ac{CC} from $0.52$ to $1.0$, and score them on the same codecs (\cref{fig:sweep}).
Values in this paragraph are relative to the unmasked codec, as in
\cref{tab:bd_main}; \cref{fig:sweep} shows the same points against the common
HEVC anchor. On the learned codec the gain is flat down to
\ac{CC}${}\approx{}0.93$, where a lightly blurred map ($\sigma=32$\,px) even
edges out the \ac{GT} map ($-11.1\%$ vs $-10.4\%$), and is still $90\%$ of the
oracle's at \ac{CC}${}\approx{}0.8$. The real predictors lie below this knee:
the remaining gap to the oracle is real, but the $38\times$ costlier predictor
does not close it. Below the knee the gain depends on the kind of error: at
\ac{CC}${}\approx{}0.67$ the prior blend keeps $-7.6\%$, like the real
predictors, whereas blur keeps $-3.6\%$ and loses its saving entirely at
\ac{CC}${}=0.52$, so \ac{CC} alone does not determine the coding gain. On HEVC
the gain grows almost linearly with \ac{CC}, with no plateau.
We attribute the slight advantage of the lightly blurred map to softer transitions
between salient and non-salient regions, which the steep shifted sigmoid of \cref{eq:gt_sat} otherwise sharpens;
Ma \textit{et al.}~\cite[Eq.~(2)]{roi_salmap_blur} likewise smooth their \ac{ROI} map.
This suggests further gain from tuning the mask generation itself.

\section{Conclusion and Future Work}
\label{sec:conclusion}
Increasing saliency-prediction accuracy yields progressively smaller coding gains for 360-degree saliency-guided compression: on the learned codec, a mid-accuracy,
near-real-time predictor captures about three quarters of the oracle's
bit-rate saving ($74\%$), a lightweight real-time predictor (52 FPS) still
captures roughly two thirds ($65\%$), while a far heavier, more accurate
predictor adds nothing measurable; the same ordering holds on HEVC. A
synthetic accuracy sweep places the learned codec's knee near
\ac{CC}${}\approx{}0.8$, above all current predictors, whereas HEVC keeps
gaining up to the oracle.
The practical takeaway for
latency-constrained 360-degree streaming (immersive teleoperation included): a lightweight predictor already delivers most of the attainable coding gain,
so the saliency stage fits comfortably within the latency constraint,
and compute is better invested in the codec itself than in pursuing saliency-metric state of the art.
The 25-\ac{ODI} test split and the dataset shift relative to the 2017
baseline are the fixed comparison context, identical for every method;
the study is limited to image-only scope.
Future work extends the study to 360-degree video within the latency constraint and
to lightweight saliency proxies built from real-time detectors.

\section*{Acknowledgment}
This research was in part conducted by AIST Researchers under FRONTia, a
Japanese national program led by METI and NEDO. This work was supported by
AIST policy-based budget project R\&D on Generative AI Foundation Models
for the Physical Domain.


\end{document}